\documentclass[%
 reprint,
superscriptaddress,
groupedaddress,
showkeys,
 amsmath,amssymb,
 aps,
floatfix,
]{revtex4-2}
\usepackage{mathrsfs}
\usepackage{amsmath}
\usepackage{bbold}
\usepackage{graphicx}
\usepackage{dcolumn}
\usepackage{bm}
\usepackage{upgreek}

\begin{document}

\preprint{APS/123-QED}

\title{Experimental realization of an active time-modulated acoustic circulator}

\author{Matthieu MALL\'EJAC}
\email{matthieu.mallejac@epfl.ch}
\author{Romain FLEURY}%
 \email{romain.fleury@epfl.ch}
 
\affiliation{%
 Laboratory of Wave Engineering, \'Ecole Polytechnique F\'ed\'erale de Lausanne, Switzerland.
}%

\date{\today}

\begin{abstract}
Reciprocity is one of the fundamental characteristics of wave propagation in linear time-invariant media with preserved time-reversal symmetry. Breaking reciprocity opens the way to numerous applications in the fields of phononics and photonics, as it allows the unidirectional transport of information and energy carried by waves. In acoustics, achieving non-reciprocal behavior remains a challenge, for which time-varying media are one of the solutions. Here, we design and experimentally demonstrate a three-port non-reciprocal acoustic scatterer that behaves as a circulator for audible sound, by actively modulating the effective mass of the acoustic membranes over time. We discuss the conception and experimental validation of such an acoustic circulator, implemented with actively controlled loudspeakers, in the realm of audible and airborne acoustics, and demonstrate its good performance in different scenarios.
\end{abstract}

\maketitle

\section{Introduction}
\label{sec1}

     The integration of time modulation as a new degree of freedom in the design of engineered systems has generated significant interest and led to the emergence of new paradigms in wave manipulation \citep{galiffi_photonics_2022}. Time-varying media have indeed demonstrated remarkable capabilities in enabling various wave phenomena, including unidirectional and parametric amplification  \citep{cullen_travelling-wave_1958, koutserimpas_nonreciprocal_2018, koutserimpas_parametric_2018, song_direction_2019, li_nonreciprocal_2019, shen_nonreciprocal_2019, shen_nonreciprocal_2019-1, zhu_tunable_2020, zhu_non-reciprocal_2020, wen_unidirectional_2022}, frequency conversion \citep{lee_linear_2018,wen_unidirectional_2022,zhao_programmable_2019}, and the induction of strong non-linearities \citep{shan_publisher_2022,mukherjee_observation_2020}. This dynamic modulation approach therefore represents a versatile and powerful tool for tailoring wave propagation.
    
    In particular, time modulation holds great promise as an effective means of breaking the reciprocal nature of wave propagation \citep{sounas_nonreciprocal_2017}. In linear systems, achieving strong non-reciprocity is in fact a fundamental challenge as reciprocity is closely linked to the time-reversal symmetry property of the wave equation \citep{onsager_reciprocal_1931, onsager_reciprocal_1931b, casimir_on_1945}, although it offers new prospects for numerous applications in different fields, from electromagnetism \citep{caloz_electromagnetic_2018} to acoustics and mechanics \citep{wang_observation_2018, verhagen_optomechanical_2017,nassar_nonreciprocity_2020,rasmussen_acoustic_2021, zhu_janus_2021, chen_efficient_2021, chen_sound_2023}.
    One of the key demonstrations of such a strong nonreciprocal propagation is the so-called circulator, which allows the steering of a wave in a specific direction while isolating in the other. Microwave circulators based on the magnetic biasing of a ferromagnetic atom are widely used in electrical engineering for example to isolate a source from unwanted reflections, allow for full-duplex communications, or create topologically protected communication paths and edge states \citep{zhang_superior_2021, zhang_anomalous_2023}. While electromagnetic waves can efficiently exploit this magnetoacoustic effect to achieve non-reciprocity, mechanical waves are more weakly influenced by magnetic fields, hence the importance of the extension to non-magnetic devices, which is still in the research and development stage.  
    Most of the magnetic free-circulator designs proposed to date \citep{estep_magnetic-free_2014,fleury_sound_2014,fleury_subwavelength_2015, pedergnana2024synchronizationbased} rely on a central cavity (cylindrical or ring) connected to three equispaced external ports, hosting two azimuthal counterpropagative modes.  Under the action of an external synthetic angular-momentum bias, the degeneracy is lifted creating constructive and destructive interferences responsible for the strong non-reciprocal response. 
    
    In the realm of airborne acoustics, the concept of acoustic circulators has also garnered significant attention as a means of controlling wave propagation without the need for magnetic biasing. Inspired by the Zeeman mode-splitting effect observed in electromagnetic systems, researchers have explored the introduction of slow circulating flows into resonant cavities \citep{fleury_sound_2014, ding_experimental_2019, pedergnana2024synchronizationbased} to induce non-reciprocal behavior in acoustic waves. Analytical predictions and numerical simulations have further suggested the feasibility of constructing ultrasonic circulators using time-varying cavities filled with a medium whose bulk modulus is periodically modulated, effectively imparting angular momentum to the acoustic waves \citep{fleury_subwavelength_2015}. While related techniques have been experimentally validated in the context of topological elastodynamics \citep{darabi_reconfigurable_2020}, the experimental realization of airborne acoustic circulators based on this dynamical approach remains unreported.

    In this work, we address this gap by presenting the design and experimental demonstration of a subwavelength airborne and audible acoustic circulator. Our approach relies on three actively controlled loudspeaker membranes, the specific impedance of each being synthetically periodically modulated in time at a frequency significantly lower than that of the acoustic wave. Through a combination of theoretical analysis, numerical simulations, and experimental validation, we showcase the effectiveness of our time-modulated circulator in achieving non-reciprocal wave propagation in airborne acoustic environments.

\section{ Time-modulated acoustic resonators, design of a circulator}
    \label{sec:design}
\begin{figure*}
        \centering
        \includegraphics[width = 0.9\textwidth]{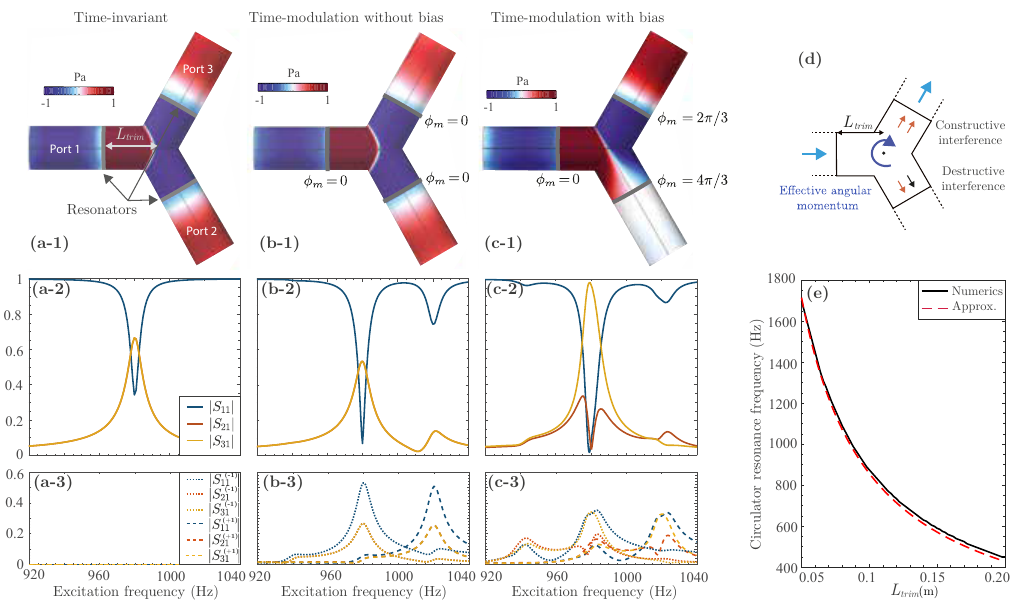}
            \caption{(Color online) \textbf{Numerical analysis of a triport waveguide system terminated by three identical one-degree of freedom resonators} in different scenarios: \textbf{(a)} time-invariant, \textbf{(b)} time-modulated but no bias (no modulation phase difference), and \textbf{(c)} time-modulated with bias (modulation phase $\phi_1 = 0$, $\phi_2 =2\pi/3$, and $\phi_3 = 4\pi/3$ assigned to each resonator). \textbf{(.-1)} Pressure distribution,  \textbf{(.-2)} reflection $|S_{11}|$, transmission to port 2 $|S_{21}|$, and transmission to port 3 $|S_{31}|$ for the fundamental frequency and \textbf{(.-3)}  the first Floquet harmonic $\omega\pm \omega_m$. \textbf{(d)} Schematic of the biased circulator, with the clockwise effective angular momentum and the constructive/destructive interference highlighted. \textbf{(e)} Evolution of the resonance frequency of the circulator with the trimmer length for given resonators: FDTD numerical results (solid line) and first-order analytical approximation (dashed line).  }
        \label{fig:concept}
    \end{figure*}

    First, we explore the design of such an acoustic time-modulated circulator. As an alternative to the ring cavity geometry proposed in the literature, in this work, we consider a circulator consisting of a triport "Y" type waveguide with 120° rotational symmetry coupled in series with three time-modulated acoustic resonators placed at each of the input ports, as shown in Fig. \ref{fig:concept} (a-1). 

    To gain a better understanding of its behavior and the interplay between each parameter of the design, and to remain general at this stage, we will first examine a simplified system where the acoustic resonators are simply replaced by a first-order differential equation, i.e. an ideal mass-spring equivalent system, for which we can vary one of the parameters with time.

        \subsection{Time-modulated mass-spring-damping equivalent resonator}
        \label{sec:num_eq_circuit}
            In Figures~\ref{fig:concept} (a),(b), and (c) respectively, the triport scattering behavior is studied numerically using the commercial software COMSOL Multiphysics for three different configurations: (i) with time-invariant resonators,  (ii) with time-modulated resonators but without momentum bias, and (iii) with time-modulation and momentum bias. In each case, an incident monochromatic wave is impinging the circulator from port 1,  and we characterize the transmission at ports 2 (red line) and 3 (yellow line) as well as the reflection at port 1 (blue line). 
            
            In the time-invariant and lossless case, propagation is reciprocal and the acoustic energy is therefore equally split between the three ports of the circulator,  giving rise to an equal transmission to ports 2 and 3 of $|S_{21}|=|S_{31}| = 2/3$ and a reflection to port 1 of $|S_{11}|=1/3$ as depicted by both the pressure distribution in Fig. \ref{fig:concept} (a-1) and the evolution of the scattering parameters with frequency in Fig. \ref{fig:concept} (a-2). 
            
            Applying a cosine modulation to one of the resonator parameters (here the capacitance $C_r$) to periodically modulate its resonance frequency in time is not sufficient to induce the desired non-reciprocal behavior, as evidenced in Fig.~\ref{fig:concept} (b).  However, although the transmitted energy is still split equally between ports 2 and 3, the transmission and reflection are reduced to 0.5 and 0.1 respectively as a result of the time modulation and the associated energy transfer to the Floquet harmonics (see Fig.~\ref{fig:concept} (b-3)), i.e. energy transfer to $f= f_0\pm n f_m$ \citep{mallejac_scattering_2023}.
          
            Non-reciprocal propagation is only triggered by introducing a momentum bias into the modulation. In Figure~\ref{fig:concept} (c), the modulation is applied in a rotating fashion by dephasing the modulation on each resonator by $\Delta \phi_m = 2\pi/3$. The effective angular momentum generated by this dynamic modulation breaks both time invariance and time reversal symmetry and allows non-reciprocal control of wave propagation. As a result, constructive interference allows perfect transmission at port 3, while destructive interference prevents the incident wave from being transmitted to port 2.  It is worth noting here that an inversion of the dephasing bias would instead result in a clockwise rotation and as a consequence in a perfect transmission at port 2 and a zero transmission at port 3.
        
        \subsection{Impact of the cavity length}
        \label{sec:impact_length}
            Since the circulation behavior is based on an interference phenomenon, the geometry of the cavity is an important characteristic of the design.
            Indeed, the operating frequency range of the device is mainly determined by the length $L_{trim}$ of the three interconnected waveguides forming the Y-type waveguide, and then, as a second-order correction, by the mechanical properties of the resonators.
            Figure \ref{fig:concept}(d) shows the evolution of the operating frequency $f_r$ with the cavity length. 
        
            As already mentioned, at the resonance frequency of the circulator, destructive interference occurs between the direct propagation path from port 1 to port 3, i.e. a propagation distance of $L_1=2L_{trim}$, and the propagation path from port 1 to port 3 after a reflection at port 2, i.e. a propagation distance of $L_2=4L_{trim}$, which must be out of phase with the direct path ($\Delta \phi = \pi$). The evolution curve derived from the numerical model (solid line) can thus be well approximated by an analytical first-order curve (dashed red line) obtained by solving
            \begin{equation}
                \Delta \phi = \omega_r/c_0 \left(L_{2}-L_{1}\right) = \pi,
            \end{equation}
            giving, $f_r = c_0/4L_{trim}$.
        
            The length of the cavity must therefore be chosen carefully. The operating frequency range should be sufficiently distant from the resonance frequency of each individual resonator, to prevent coupling and interference between the two resonances, that might alter the highly non-reciprocal behavior.

        \subsection{Choice of the resonators: actively controlled loudspeakers}
        \label{sec:res_choice}
            Among the various choices of individual resonators that could make up the acoustic version of the time-modulated circulator, we decided to use three electrodynamic loudspeakers. In the low-frequency limit, a loudspeaker can be approximated by a lumped element model consisting of a mechanical mass (inductance) $ M_{\textrm{ms}}$, a resistance $ R_{\textrm{ms}}$ and a capacitance $ C_{\textrm{ms}}$.  The use of loudspeakers has two main advantages: they can be easily modeled (see Section 2 of Material and Methods) and they can also be actively controlled.  With active control, the response of the resonators can be synthetically modified from their natural behavior, for example by changing their resonant frequency, stiffness or impedance \citep{rivet_broadband_2017, koutserimpas_active_2019,guo_improving_2020, guo_pid-like_2022}, thus offering great versatility for the implementation of time-modulated elements \citep{mallejac_scattering_2023}.
            
            Here we periodically modulate the specific impedance $Z_s$ of each loudspeaker in time by applying a broadband active control so that they respond with a target impedance $Z_{\textrm{targ}}(t)$. The control scheme consists of first measuring the pressure difference $\Delta p$ across the diaphragm and then applying a feedforward loop that assigns a given current $i(t)$ to the loudspeaker on the basis of a given control law $\Theta$ \citep{mallejac_scattering_2023}
            \begin{equation}
                        i = \Theta \Delta p = \frac{S_d}{Bl}\left(1-\frac{Z_{s}}{Z_{\textrm{targ}}(t)}\right)\Delta p,
                        \label{eq:control_law}
            \end{equation}
            where $S_d$ is the effective surface of the diaphragm,  $Bl$ is the force factor, $Z_{s}$ is the loudspeaker specific impedance, and $Z_{targ}$ the specific impedance we want to synthesize, which can depend on time.      
            
            The bias is introduced in our study by synthetically varying periodically in time the mechanical mass of the loudspeaker membranes.
            The target impedance is therefore set as
            \begin{equation}
            Z_{targ} = R_{\textrm{ms}} +\left[ \textrm{i}\omega C_{\textrm{ms}} \right]^{-1}+\textrm{i} \omega M_{\textrm{ms}}\left(1+A_m\cos\left(\omega_m t + \phi_{m}\right)\right),
            \end{equation}
            where $\omega_m$, $A_m$, and $\phi_m$ are the modulation circular frequency,  depth, and phase respectively. 
            
\begin{figure}
            \centering
            \includegraphics{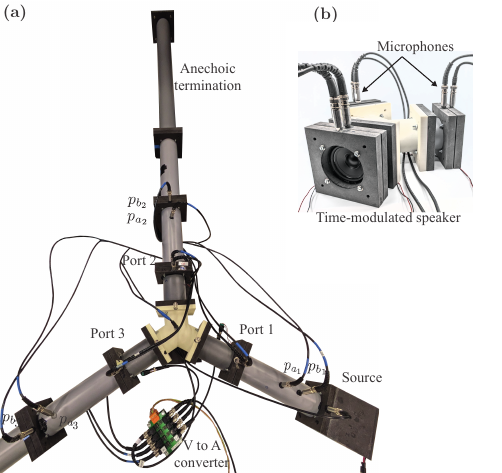}
                     \caption{(Color online) \textbf{Experimental set-up.} Photographies the experimental set-up (a) and closed-up on the actively controlled loudspeaker (b).}
            \label{fig:set-up}
            \end{figure}

            Such an active control scheme requires a good response from the loudspeaker and is therefore mainly efficient and stable around its resonant frequency. We therefore have to compromise between these two constraints, i.e., the control stability constraint and the mode decoupling constraint. We first select the loudspeakers to be controlled and then fix the operating frequency range of the circulator by adequately choosing the length of the three interconnected waveguides forming the Y-type waveguide (see Fig.~\ref{fig:set-up}), fixed here at $L_{trim} = 16$ cm, so that the operating frequency $f = 569$ Hz is sufficiently far from the loudspeaker resonance $f = 170$ Hz, while staying in the control stability range.  It is worth noting that these constraints can also be leveraged for a fixed cavity geometry by adapting the natural mechanical properties of the loudspeaker through active control.
            
\begin{figure*}
        \centering
        \includegraphics[width = 0.9\textwidth]{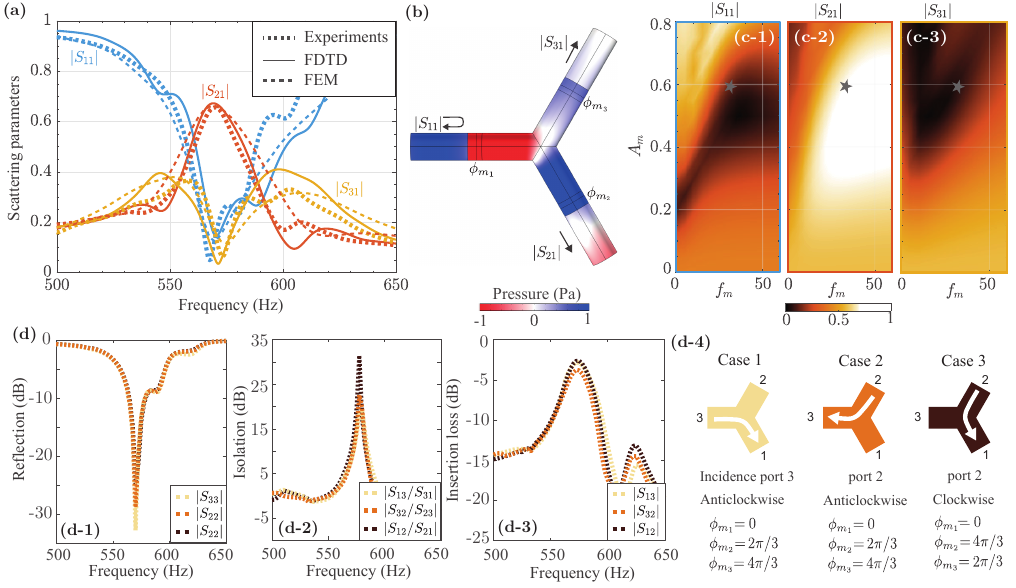}
        \caption{(Color online) \textbf{Experimental results.} \textbf{(a)} Circulator scattering parameters: $S_{11}$ (blue), $S_{21}$ (red), and $S_{31}$ (yellow) obtained from measurements (points), Finite Difference Time Domain (solid lines), and Finite Element Method (dashed lines). 
        \textbf{(b)} Pressure field distribution of the optimized circulator obtained from full-wave simulation. \textbf{(c)} Dependency of $\mathbf{S}$ with the modulation frequency $f_m$ and depth $A_m$ :  $|S_{11}|$ \textbf{(c-1)}, $|S_{21}|$ \textbf{(c-2)}, and $|S_{31}|$ \textbf{(c-3)} (FDTD).  (\textbf{d}) Comparison of the reflection $20\log_{10}(|S_{ii}|)$ \textbf{(d-1)}, Isolation $20\log_{10}(|S_{ji}/S_{li}|)$ \textbf{(d-2)}, and Insertion loss $20\log_{10}(|S_{li}|)$ \textbf{(d-3)} for an incidence at port 1 and an anticlockwise bias (brown, case 1), for an incidence at port 1 but a clockwise bias (orange, case 2), and for an incidence at port 3 and an anticlockwise bias (light beige, case 3).}
        \label{fig:results}
    \end{figure*}

            On the other hand, the circulator performance is also closely related to the choice of modulation amplitude and frequency. The optimal configuration is found by exploring in parameter space ($A_m$, $f_m$) the evolution of the scattering parameters at each port for an incidence at port 1, i.e. $|S_{11}|$, $|S_{21}|$ and $|S_{31}|$, as shown by the simulated maps in Figs. \ref{fig:results} (c) obtained from finite-difference time-domain simulations (see Methods).  The optimal modulation parameters, highlighted by the grey stars in Fig. \ref{fig:results} (c), $A_m = 0.6$ and $f_m=31$ Hz, are found by compromising between low reflection (Fig. \ref{fig:results} (c-1)), high transmission at port 2 (Fig. \ref{fig:results} (c-2)) and low transmission at port 3 (Fig. \ref{fig:results} (c-3)).
            
\section{\label{sec:texp}Results and discussion}
\label{sec:res}
     Once the circulator had been designed numerically, we set up the experimental apparatus shown in Figure \ref{fig:set-up} (a-b). To evaluate its performance experimentally, we connected the ports of the circulator to external waveguides of the same cross-section and instrumented them with pairs of microphones to extract the acoustic scattering matrix of the system (see Methods). Prior to any measurements, we experimentally characterized the mechanical properties of the loudspeakers (see Methods) and carried out an experimental exploration of the parameter space ($A_m$, $\omega_m$) similar to the numerical one (shown in Fig. \ref{fig:results} (c)). For a modulation frequency of $f_m = 31$ Hz, the optimal modulation depth is $A_m = 0.8$. The slight deviations from the optimal value observed in the numerical FDTD method could be explained by several factors. Indeed, the numerical results are based on several approximations: the main ones being that the viscothermal losses are not taken into account, the loudspeakers are modeled as single-degree-of-freedom resonators, and the complex coupling (acoustic and magnetic) between the three loudspeakers is not accounted for.
     
     In Figure \ref{fig:results} (a) we compare the scattering parameters obtained on the three-port system using a modulation depth of 0.8 for the experimental results (thick dotted line) and a modulation depth of 0.6 for the FDTD simulations (solid line). We also cross-validated the results with a finite element model (see Methods), represented here by the dashed line. The results in Fig. \ref{fig:results} (a) together with the full-wave pressure field in Fig. \ref{fig:results} (b) clearly show the good circulation behavior of the system under periodic cosine modulation of frequency 31 Hz, with a transmission at port 2 of $|S_{21}|=0.68$ and a transmission at port 3 of $|S_{21}|=0.05$. 
     It should be noted that the experimental result is in close agreement with the FEM and FDTD simulations, despite the approximations mentioned above.

    To go further, we evaluate the performance of a circulator for an incidence at any port $i$ and transmission at port $j$ using the three main classical metrics: (1) the reflection coefficient $R=20\log\left(|S_{ii}|\right)$, which must be as low as possible, (2) the isolation $IS=20\log\left(|S_{ji}/S_{ij}|\right)$, which quantifies the circulator's ability to transmit in a single direction and should be as high as possible, and (3) the forward insertion loss $IL = 20\log\left(|S_{ji}|\right)$, which characterizes the amplitude loss during transmission and should therefore be as close to zero as possible. \\
    
    We tested three different cases to cover all possibilities and fully characterize the designed device. 
    First, we generated an incident pressure wave from port 3 and applied a counterclockwise bias ($\Phi_{m_1}=0$, $\Phi_{m_2}=2\pi/3$, $\Phi_{m_3}=4\pi/3$) to direct the wave towards port 1, as shown by the beige lines in the experimental results in Figs. \ref{fig:results} (d). We then tested a transmission from port 2 to port 3 with the same bias, as shown by the orange lines. Finally, we reversed the bias to obtain a clockwise rotation ($\Phi_{m_1}=0$, $\Phi_{m_2}=4\pi/3$, $\Phi_{m_3}=2\pi/3$), with a transmission from port 2 to port 1 as the final validation, shown in brown.
    
    The results obtained in all three tests demonstrate the good performance of our easy-to-implement airborne circulator, with a reflection of over -25 dB, an isolation of over 20 dB and a low insertion loss of just a few decibels for each configuration.

\section{Conclusions}
\label{sec:conclu}
    In conclusion, we have designed and experimentally demonstrated for the first time an airborne linear circulator based on time modulation. By actively controlling and dynamically synthetically modulating the loudspeaker mass, we were able to demonstrate non-reciprocal acoustic wave circulation with -25 dB reflection and 20 dB isolation. Although the inevitable dissipation prevents perfect forward transmission, we observed a low insertion loss of around 4 dB. The good performance of the circulator paves the way for its use in airborne Floquet topological insulators, for which the critical criterion of a reflection of less than 1/3 is met \citep{zhang_superior_2021}. The main advantage of the proposed device is its scalability and versatility, thanks to the use of active control. One of the main prospects would therefore be to miniaturize the three-port scatterer and use analog rather than digital active control to facilitate its integration into large networks.

\section{Acknowledgements}
The authors acknowledge Dr. Z. Zhang for fruitful discussions.
\appendix

\section*{\label{app:model}Material and methods}
    
    \subsection{Scattering measurement}
    The experimental set-up shown in Fig.\ref{fig:set-up} (a-b) consists of a source loudspeaker (Monacor SPX30 M, 3 inches) and three actively controlled loudspeakers (Visaton FR8 TA, full range, 3 inches) facing the output ports (see zoom in inset (a)) and positioned inside a 7.18 cm diameter air-filled cylindrical rigid waveguide. To consider only the propagation of plane waves, we take care to work only below the 1st cutoff frequency of the waveguide ($f_c = 1.8412 c_0/2\pi a = 2800$ Hz).

    Three pairs of microphones (PCB Piezotronics 130F20, 1/4 inches) are used to distinguish incoming waves (superscript +) from outgoing waves (superscript -) at each port ($l = 1,2$ or $3$)
    \begin{equation}
          \begin{bmatrix}
                p_{l}^{+} & p_{l}^{-}\\
            \end{bmatrix}
            =
             \begin{bmatrix}
                p_{a_l} & p_{b_l}\\
            \end{bmatrix}\cdot
            \begin{bmatrix}
                \textrm{e}^{-\textrm{i}k z_a} & \textrm{e}^{-\textrm{i}k z_b}\\
                \textrm{e}^{\textrm{i}k z_a} & \textrm{e}^{\textrm{i}k z_b}\\
            \end{bmatrix}^{-1},
            \label{eq:incident_ref_mat}
        \end{equation}    
        where $p_{a_l}$ and $p_{b_l}$ are the pressure measured by each pair of microphones at a distance $z_a=15$ cm and $z_b=20$ cm from the input of the circulator (see Fig.\ref{fig:set-up}(c))
            
        The measurements are performed in a multi-step procedure where each configuration is tested separately. The source is placed on ports 1, 2 or 3 in turn and the other two ports are connected to anechoic terminations to avoid reflections and to facilitate the extraction of the scattering parameters. The system is excited by a stepped swept sine wave with a step duration of 4 s. The active control scheme, excitation signal generation, and data acquisition are performed using a Speed-goat Performance real-time controller (I/O 135) controlled by the MATLAB/SIMULINK xPC target environment.
        
        The scattering parameters are then derived as follows
        \begin{align}
            S_{ll} & = p_{l}^-/p_l^+, \\
            S_{kl} & = p_{k}^-/p_l^+.
        \end{align}
        where $S_{ll}$ and $S_{kl}$ are the reflection at port $l$, and the transmission  at port $k$ for an incidence at port $l$ respectively,

    \subsection{Modeling of the circulator and control law}
    In the low-frequency limit, an electrodynamic loudspeaker can be approximated by a lumped element model consisting of a mechanical mass (inductance) $ M_{\textrm{ms}}$, a resistance $ R_{\textrm{ms}}$ and a capacitance $ C_{\textrm{ms}}$.
    
\begin{figure}
    \centering
    \includegraphics{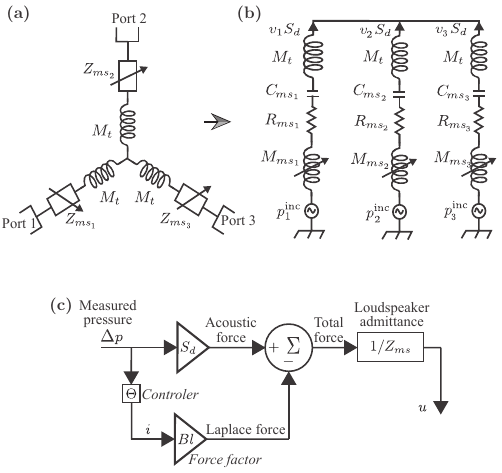}
             \caption{ \textbf{Lumped model and control law.} Lumped-model representation of the circulator (a) and (b). Active control scheme applied to each loudspeaker (c).}
    \label{fig:appdset-up}
    \end{figure}

    The overall circuit of the designed circulator is therefore a parallel assembly of the three loudspeaker circuits connected in series, with an additional mass $M_t$ corresponding to each part of the waveguide forming the Y-type waveguide, as shown in Fig. \ref{fig:appdset-up} (a,b).

        The time-modulated resonators are actively controlled electrodynamic loudspeakers (Visaton FR8 TA, 3 inches) enclosed in a cylindrical waveguide and instrumented with two ICP microphones (PCB 130F20, 1/4 inches) placed on either side of the loudspeaker diaphragm. The system is controlled by a Speed-Goat Performance real-time controller (I/O 135). The controller's output voltage is converted by a homemade voltage-to-current converter (0.10 A/V) based on a Howland pump circuit and fed to the controlled loudspeakers. 

        Some additional settings were chosen to ensure good circulation. The performance of a circulator is indeed intrinsically dependent on the visco-thermal and mechanical losses in the system. While the former are fixed for a given waveguide, the latter can be modified by active control. To increase the Q-factor of the resonances and thus the performance of the circulator, we synthetically reduced the mechanical resistance of each loudspeaker by $R_{ms}/10$. In addition, to ensure that all the loudspeakers respond identically, we have also synthetically assigned to each loudspeaker the mean values of the mechanical parameters $R_{ms_m}$, $M_{ms_m}$ and $C_{ms_m}$ of the three loudspeakers by applying an active control scheme alongside the periodic modulation scheme.

        The complete control law, depicted in Fig.~\ref{fig:appdset-up}(c) therefore reads as

        \begin{widetext}
            \begin{equation}
            i(t) = \Theta(t)\Delta p = \frac{S_d}{Bl} \left( \frac{\textrm{i}\omega M_{ms} (\mu_m(t)-1)+R_{ms}(\mu_r-1)+[\textrm{i}\omega C_{ms}]^{-1}(\mu_c-1)}{\textrm{i}\omega M_{ms} \mu_m(t) + R_{ms}\mu_r+[\textrm{i}\omega C_{ms}]^{-1}(\mu_c-1)}\right),
            \end{equation}
        \end{widetext}

        where 
        \begin{align*}
            \mu_r&=R_{ms_m}/(10 R_{ms}), \\
            \mu_c&=C_{ms_m}/C_{ms},  \\
            \mu_m&=M_{ms_m}\left[\left(1+A_m\cos(\omega_m t+\phi_m\right)\right]/M_{ms}
        \end{align*}

    \subsection{Numerical modelling}
    Two numerical models have been developed to design the circulator and cross-check the experimental results, based on the finite element method and finite difference time domain. 
    
   \subsubsection{Numerical FDTD model}
        The FDTD numerical results are obtained using SIMULINK modeling of the entire experimental setup, based on a finite difference time domain approach using a time step of $50$ microseconds. The scattering parameters can be extracted using two probes at each port. The results obtained from the extraction procedure performed as in the experimental setup are consistent with those obtained from the direct access to the reflected and transmitted pressures allowed by the simulation.

    \subsubsection{Numerical FEM model}    
    The numerical FEM experiment is performed using the frequency domain solver of the commercial software COMSOL Multiphysics, following the methodology proposed in \citep{fleury_subwavelength_2015}. The time-modulated loudspeaker is modeled as an impedance
    \begin{align}
        Z &= Z_{ms}(\omega)+A_m \textrm{i}\omega M_{ms}/S_d \cos(\omega_m t+\phi_m)\\
        &=Z_{ms}(\omega)+\delta Z(\omega)(e^{\textrm{i}\omega_m t}e^{\textrm{i}\phi_m}+e^{\textrm{-i}\omega_m t}e^{\textrm{-i}\phi_m}).
    \end{align}
    
    An impedance condition is implemented as follows
    \begin{equation}
        -\mathbf{n}\frac{\mathbf{\nabla} p}{\rho_0} = \Delta p \frac{-\textrm{i}\omega}{Z}=-\mathbf{v}.\mathbf{n}\textrm{i}\omega.
    \end{equation}

    Expanding $\Delta p$ and $v$ (normal particle velocity) in Fourier series, we end up after some algebra to 
    \begin{align}
        \Delta p_n & =Z(\omega_n)v_n + \delta Z(\omega_n)\left(v_{n-1}e^{-\textrm{i}\phi_m} + v_{n+1}e^{\textrm{i}\phi_m}+...\right), \\
        & \textnormal{or equivalently}  \nonumber \\
        v_n & =\frac{p_{f_n}-\delta(\omega_n)\left(v_{n-1}e^{-\textrm{i}\phi_m} + v_{n+1}e^{\textrm{i}\phi_m}+...\right)}{Z(\omega_n)},
        \label{weak}
    \end{align}
    where $\omega_n = \omega+n \omega_m$.

    We then put eq.~\eqref{weak} into weak forms and solve for $n = (-2,-1,0,1,2)$ simultaneously for any incident frequency $\omega$.

    \subsection{Characterization of the resonator to control}
    A fitting procedure of the transfer function of the loudspeaker terminated by an open circuit, a short circuit, or a load $R=100.8$ $\Omega$, allows to obtain the following Thiele and Small parameters (averaged over the three loudspeakers) \citep{small1972closed-box}: $Bl_m = 2$ T·m, $M_{ms_m}= 1.2$ g, $C_{ms_m}= 0.7$ mm/N, and $R_{ms_m}= 0.56$ N.s/m. The natural resonance of the speaker occurs close to $170$ Hz. 
                                                                                                 
\end{document}